\documentclass[conference]{IEEEtran}
\IEEEoverridecommandlockouts
\usepackage{layout}
\usepackage{graphicx}
\usepackage{amssymb}
\usepackage{amsmath}
\usepackage{cite}
\usepackage{mathrsfs}
\usepackage[displaymath,mathlines]{lineno}
\usepackage{xcolor}
\usepackage{multirow}
\usepackage{algpseudocode}
\usepackage{algorithm,algpseudocode}
\usepackage{algorithmicx}
\usepackage{pbox}
\usepackage{multicol}
\usepackage{lipsum}
\usepackage{amsthm}
\usepackage{lipsum}
\usepackage{epstopdf}
\usepackage{makecell}
\usepackage{diagbox}
\usepackage{enumitem}

\newcommand{\doublewidetilde}[1]{{%
		\mathpalette\double@widetilde{#1}%
	}}
	
\makeatletter

\def\BState{\State\hskip-\ALG@thistlm}
\makeatother

\newtheorem{theorem}{Theorem}
\newtheorem{lemma}{Lemma}

\newtheorem{remark}{Remark}
\newtheorem{assumption}{Assumption}

\newcounter{eqnback}
\newcounter{eqncnt}

\begin{document}
%
\title{Sum Spectral Efficiency Maximization in Massive MIMO Systems: Benefits from Deep Learning}

\author{\IEEEauthorblockN{Trinh Van Chien, Emil Bj\"{o}rnson, and Erik G. Larsson}
	\IEEEauthorblockA{Department of Electrical
		Engineering (ISY), Link\"{o}ping University, SE-581 83 Link\"{o}ping, Sweden\\
		\{trinh.van.chien, emil.bjornson, erik.g.larsson\}@liu.se}
	\thanks{This paper was supported by the European Union's Horizon 2020 research and innovation programme under grant agreement No 641985 (5Gwireless). It was also supported by ELLIIT and CENIIT. The authors would like to thank Thuong Nguyen Canh for the fruitful discussions.}
}

\maketitle

\begin{abstract}
This paper investigates the joint data and pilot power optimization for maximum sum spectral efficiency (SE) in multi-cell Massive MIMO systems, which is a non-convex problem. We first propose a new optimization algorithm, inspired by the weighted minimum mean square error (MMSE) approach, to obtain a stationary point in polynomial time. We then use this algorithm together with deep learning to train a convolutional neural network to perform the joint data and pilot power control in sub-millisecond runtime, making it suitable for online optimization in real multi-cell Massive MIMO systems. The numerical result demonstrates that the solution obtained by the neural network is $1\%$ less than the stationary point for four-cell systems, while the sum SE loss is $2\%$ in a nine-cell system.
\end{abstract}

\IEEEpeerreviewmaketitle

\section{Introduction}
Massive MIMO (multiple-input multiple-output) is an emerging technology for cellular networks, where each base station (BS) is equipped with hundreds of antennas and  able to spatially multiplex tens of users on the same time-frequency resource \cite{Marzetta2010a,massivemimobook}. Massive MIMO  increases the spectral efficiency (SE) by the orders of magnitude compared to conventional cellular networks \cite{massivemimobook}. In case of i.i.d.~Rayleigh fading channels, Massive MIMO achieves nearly optimal SE using linear detection techniques such as maximum ratio combining (MRC) or zero-forcing \cite{Bjornson2016b}. Since there are many users that  interfere with each other in Massive MIMO, power control is essential to achieve high SE \cite{Bjornson2016b}. The channels are estimated from uplink pilots, thus both  the power of the uplink pilot and data signals can be optimized \cite{Victor2017a}. An important feature of Massive MIMO is that the SE expressions can be obtained in closed form, making it possible to efficiently optimize the powers based on the large-scale fading coefficients instead of the small-scale fading realizations, so that the power control exploits diversity against fading instead of inverting the fading.

Deep learning (DL) has become a popular way to solve problems in a data-driven fashion and has shown great performance in applications such as image restoration and pattern recognition. DL is capable of providing simple and efficient solutions, given that the complicated design and training phases have been successful. DL has recently been applied to power-control problems in wireless communications.
The authors of \cite{sun2017learning} construct a deep neural network for optimizing the sum SE in a system serving a few tens of users. The neural network structure is reused in \cite{zappone2018model} for solving an energy-efficiency problem. These works utilize classical  fully connected feed-forward networks with many layers. When these networks are used, instead of directly solving the original problems, there is a substantial performance loss: it is $5\%$ for a system serving $10$ users in \cite{zappone2018model} and  $16\%$ in case of $30$ users in \cite{sun2017learning}. 
Moreover, the optimization in these prior works is based on the instantaneous channel realizations, which requires the problems to be solved every few milliseconds and to combat bad channel realization rather than exploiting the long-term benefits of fading. The use of such methods is not scalable in Massive MIMO, where the number of channel coefficients is proportional to the number of antennas and users.

In this paper, we first formulate a joint data and pilot power control for maximizing the ergodic sum SE. The non-convexity of this problem is overcome by proposing a new algorithm, inspired by the weighted MMSE approach, that finds a stationary point instead of seeking the global optimum with exponential computational complexity. We then explore the possibility of using DL to train a neural network to solve the power control problem in sub-millisecond runtime. To this end, we model the power optimization as a supervised learning problem where the transmit powers should be predicted based on an input of  large-scale fading coefficients. Instead of using a fully-connected feed-forward network as in  \cite{sun2017learning,zappone2018model}, we use a deep convolutional neural network (CNN) and show that it achieves very high accuracy in the power prediction. Our proposed deep CNN is named PowerNet and utilizes the state-of-the-art residual structure \cite{he2016deep, zhang2018a} and dense connection \cite{huang2017}. 
The loss in SE from using PowerNet, instead of solving the power control problem directly, is only about $2\%$ in a network with $9$ cells and $10$ users per cell, while the runtime is 0.03\,ms.

\textit{Notation}: Upper (lower) bold letters are used to denote matrices (vectors). $\mathbb{E} \{ \cdot \}$ is the expectation of a random variable. The Frobenius norm denotes as $\| . \|_F$ and $\mathcal{CN}(\cdot, \cdot)$ is circularly symmetric complex Gaussian distribution. 

\section{System Model} \label{Section: System Model}
We consider the uplink of a multi-cell Massive MIMO system comprising $L$ cells, each having a BS equipped with $M$ antennas and serving $K$ users. 
Since the propagation channels vary over time and frequency, we divide the time-frequency resources into coherence intervals of $\tau_c$ symbols where the channels are static and frequency flat \cite{massivemimobook}. The channel between user~$t$ in cell~$i$ and BS~$l$ is denoted as $\mathbf{h}_{i,t}^l \in \mathbb{C}^{M}$ and follows an i.i.d.~Rayleigh fading distribution:
\begin{equation}
\mathbf{h}_{i,t}^l \sim \mathcal{CN} \left(\mathbf{0}, \beta_{i,t}^l \mathbf{I}_{M} \right),
\end{equation}
where $\beta_{i,t}^l$ is large-scale fading coefficient that models geometric pathloss and shadow fading. The distributions are known at the BSs, but the realizations are unknown and estimated independently in every coherence interval using a pilot phase.

\subsection{Uplink Pilot Transmission Phase}
We assume that a set of $K$ mutually orthogonal $K$-length pilot signals are used in the system. 
User~$k$ in each cell uses the pilot $\pmb{\psi}_{k} \in \mathbb{C}^{K}$ with $\|\pmb{\psi}_{k} \|^2 = K$. The channel estimation in cell $l$ is therefore interfered by the users in other cells, which is called pilot contamination. The received signal $\mathbf{Y}_l \in \mathbb{C}^{M \times K}$ at BS~$l$ from the pilot transmission is
\begin{equation}
\mathbf{Y}_l = \sum_{i=1 }^L \sum_{t=1}^K \sqrt{\hat{p}_{i,t}}\mathbf{h}_{i,t}^l \pmb{\psi}_{t}^H + \mathbf{N}_l,
\end{equation}
where $\mathbf{N}_l \in \mathbb{C}^{M \times K }$ is the additive noise with independent $\mathcal{CN}(0, \sigma_{\mathrm{UL}}^2)$ elements. Meanwhile, $\hat{p}_{i,t}$ is the pilot power that user~$t$ in cell~$i$ allocates to its pilot transmission. By applying MMSE estimation, a BS obtains estimates of the channels from its users and uses these when receiving the uplink data.

\subsection{Uplink Data Transmission Phase}
During the uplink data transmission phase, user~$t$ in cell~$i$ transmits data symbol $s_{i,t}$ with $\mathbb{E} \{ |s_{i,t}|^2\} =1$. The received signal $\mathbf{y}_{l} \in \mathbb{C}^{M}$ at BS~$l$ is the superposition of the signals transmitted from all users and is given by
\begin{equation}
\mathbf{y}_{l} = \sum_{i=1}^L \sum_{t=1}^K \sqrt{p_{i,t}} \mathbf{h}_{i,t} s_{i,t} + \mathbf{n}_{l},
\end{equation}
where $p_{i,t}$ is the power that user $t$ in cell $i$ allocates to the data transmission. $\mathbf{n}_{l}  \sim \mathcal{CN} \left(\mathbf{0}, \sigma_{\mathrm{UL}}^2 \mathbf{I}_{M} \right)$ is the additive noise. 
We assume that each BS uses MRC (based on MMSE channel estimates) to detect the desired signals from its users. Using the standard Massive MIMO methodology, the following closed-form lower bounds on the ergodic capacities are obtained.

\begin{lemma}[\!\!\cite{Chien2017a}, Corollary~$1$] \label{Lemma1}
If the BSs use MRC for data detection, an achievable ergodic SE of user $k$ in cell $l$ is 
\begin{equation} \label{eq:ULRate}
R_{l,k} = \left(1 -\frac{K}{\tau_c} \right) \log_2 \left( 1 + \mathrm{SINR}_{l,k}\right),
\end{equation}
where the effective SINR value of this user is
\begin{equation} \label{eq:SINRlk}
\mathrm{SINR}_{l,k} = M K p_{l,k}  \hat{p}_{l,k} (\beta_{l,k}^l)^2 / \mathit{D}_{l,k}
\end{equation} 
and the interference and noise term $\mathit{D}_{l,k}$ is given by
\begin{equation}
\begin{split}
 &\mathit{D}_{l,k} = M K  \sum\limits_{\substack{i=1, i \neq l}}^L p_{i,k} \hat{p}_{i,k} (\beta_{i,k}^l)^2  \\
 & + \left( K \sum\limits_{i=1}^L \hat{p}_{i,k} \beta_{i,k}^l + \sigma_{\mathrm{UL}}^2  \right)\left( \sum\limits_{i=1}^L \sum\limits_{t=1}^K p_{i,t} \beta_{i,t}^l + \sigma_{\mathrm{UL}}^2 \right).
\end{split}
\end{equation}
\end{lemma}
In \eqref{eq:SINRlk}, the numerator contains the array gain which is proportional to the number of antennas $M$ at the receiving BS. The first part in the denominator is the effect of the pilot contamination and it is also proportional to $M$. The remaining terms are non-coherent mutual interference and noise, which are independent of $M$ and thus are negligible when the number of antennas is very large.

\section{Joint Pilot and Data Power Control for Sum Spectral Efficiency Maximization}
\label{section:joint-optimization}

In this section, we formulate a sum SE maximization problem where the pilot and data powers are jointly optimized. This important problem has an inherent non-convex structure, different from the single-cell case in \cite{Victor2017a} which has an efficient solution. Hence, our first contribution is to derive an iterative algorithm that obtains a stationary point in polynomial time, by solving a series of convex sub-problems.

\subsection{Problem Formulation}
\setcounter{eqnback}{\value{equation}} \setcounter{equation}{14}
\begin{figure*}
	\begin{equation} \label{eq:rhohatlk}
	\hat{\rho}_{l,k}^{(n)} = \min \left( \frac{ \sqrt{M  K} \rho_{l,k}^{(n-1)} u_{l,k}^{(n)} w_{l,k}^{(n)} \beta_{l,k}^l}{(\rho_{l,k}^{(n-1)})^2 M K \sum\limits_{i=1}^L w_{i,k}^{(n)} (u_{i,k}^{(n)})^2 (\beta_{l,k}^i)^2  + K \sum\limits_{j=1}^L w_{j,k}^{(n)}  (u_{j,k}^{(n)})^2 \beta_{l,k}^j \left( \sum\limits_{i=1}^L \sum\limits_{t=1}^K (\rho_{i,t}^{(n)})^2 \beta_{i,t}^j + \sigma_{\mathrm{UL}}^2 \right) } , \sqrt{P}\right) 
	\end{equation} \vspace*{-0.3cm}
	\begin{equation} \label{eq:rholk}
	\rho_{l,k}^{(n)} = \min \left( \frac{\sqrt{MK} \hat{\rho}_{l,k}^{(n)} u_{l,k}^{(n)} w_{l,k}^{(n)} \beta_{l,k}^l }{ (\hat{\rho}_{l,k}^{(n)})^2 M K \sum\limits_{i=1}^L w_{i,k}^{(n)} (u_{i,k})^2 (\beta_{l,k}^i)^2 +  \sum\limits_{i=1}^L \sum\limits_{t=1}^K w_{i,t}^{(n)} (u_{i,t}^{(n)})^2 \beta_{l,k}^i \left( K \sum\limits_{j=1}^L (\hat{\rho}_{j,t}^{(n)})^2 \beta_{j,t}^i + \sigma_{\mathrm{UL}}^2\right) }, \sqrt{P}\right)
	\end{equation} 
	\hrule
	\vspace*{-0.45cm}
\end{figure*}
\setcounter{eqncnt}{\value{equation}}
\setcounter{equation}{\value{eqnback}}

We want to maximize the sum SE of the $LK$ users under constraints on the power per pilot and data symbol:
\begin{equation} \label{Prob:SumRate}
\begin{aligned}
& \underset{\{ \hat{p}_{l,k}, p_{l,k} \geq 0 \} }{\textrm{maximize}}
&&   \sum_{l=1}^{L} \sum_{k=1}^K R_{l,k}\\
& \,\,\textrm{subject to}
&&   \hat{p}_{l,k} \leq P, \; \forall l,k,\\
&&& p_{l,k} \leq  P, \; \forall l,k,
\end{aligned}
\end{equation}
where $P$ is the maximum power that each user can supply to its transmitted symbols. Plugging \eqref{eq:ULRate} into \eqref{Prob:SumRate} and removing the constant pre-log factor, we obtain the equivalent problem
\begin{equation} \label{Prob:SumRatev2}
\begin{aligned}
& \underset{\{ \hat{p}_{l,k}, p_{l,k} \geq 0 \} }{\textrm{maximize}}
&&   \sum_{l=1}^{L} \sum_{k=1}^K \log_2 \left(1 + \mathrm{SINR}_{l,k} \right)\\
& \,\,\textrm{subject to}
&&   \hat{p}_{l,k} \leq P, \; \forall l,k,\\
&&& p_{l,k} \leq  P, \; \forall l,k.
\end{aligned}
\end{equation}
This problem is independent of the small-scale fading due to the SINR expression in \eqref{eq:SINRlk}. Hence, its solution can be used for a long period of time, if the users are continuously active and there is no large-scale user mobility. However, in practical systems, some users are moving quickly (such that $\beta_{i,t}^{l}$ changes) and new scheduling decisions are made every few milliseconds based on the users' data traffic. It is important to be able to solve \eqref{Prob:SumRatev2} very quickly to adapt to these changes.\footnote{Note that the ergodic SE is a reasonable performance metric also in this scenario, since long codewords can span over the frequency domain and the channel hardening makes the channel after MRC almost deterministic.}

Inspired by the weighted MMSE methodology \cite{Christensen2008a}, we will now propose an iterative algorithm to find a stationary point. To this end, we define $\hat{\rho}_{l,k} = \sqrt{\hat{p}_{l,k} }$ and $\rho_{l,k} = \sqrt{p_{l,k} }$,$\forall l,k,$ as new optimization variables and the derive following theorem obtains a new problem formulation that is equivalent with \eqref{Prob:SumRatev2}.
\begin{theorem} \label{Theorem:WMMSE}
The following optimization problem is equivalent to problem \eqref{Prob:SumRate}: 
\begin{equation} \label{Prob:WMMSEv1}
\begin{aligned}
& \underset{\substack{ \{ w_{l,k} \geq 0, u_{l,k} \}, \\ \{ \hat{\rho}_{l,k}, \rho_{l,k} \geq 0 \} }}{\mathrm{minimize}}
&&   \sum_{l=1}^{L} \sum_{k=1}^K w_{l,k} e_{l,k} - \ln (w_{l,k}) \\
& \,\,\,\mathrm{subject\,to}
&&   \hat{\rho}_{l,k}^2 \leq P, \; \forall l,k,\\
&&& \rho_{l,k}^2 \leq  P, \; \forall l,k,
\end{aligned}
\end{equation}
where $e_{l,k}$ is given by
\begin{align} \notag
& e_{l,k} = MK u_{l,k}^2 \sum_{i=1}^L \rho_{i,k}^2 \hat{\rho}_{i,k}^2 (\beta_{i,k}^l)^2 - 2 \sqrt{M  K} \rho_{l,k} \hat{\rho}_{l,k} u_{l,k} \beta_{l,k}^l  \\  &+ u_{l,k}^2 \!\left( K \sum_{i=1}^L \hat{\rho}_{i,k}^2  \beta_{i,k}^l + \sigma_{\mathrm{UL}}^2 \right) \!\! \left( \sum_{i=1}^L \sum_{t=1}^K \rho_{i,t}^2 \beta_{i,t}^l + \sigma_{\mathrm{UL}}^2 \right)
\!+\!1.  \label{Prob:WMMSEv2}
\end{align}
More precisely, if $\{u_{l,k}^{\ast}, w_{l,k}^{\ast}, \hat{\rho}_{l,k}^{\ast}, \rho_{l,k}^{\ast} \}$ is a global optimum to \eqref{Prob:WMMSEv1}, then  $\{ (\hat{\rho}_{l,k}^{\ast})^2, (\rho_{l,k}^{\ast})^2 \}$ is a global optimum to \eqref{Prob:SumRate}.
\end{theorem}
\begin{IEEEproof}
The main procedure is similar to \cite{Chien2018a}, where only the data powers were optimized while the pilot powers were constant. The proof for the extension to  joint data and pilot power control can be done in two main steps. We first derive the mean square error $e_{l,k}$ for user $k$ in cell $l$, considering a single-input single-output (SISO) communication system with deterministic channels having the same SE as in Lemma~\ref{Lemma1}, where $u_{l,k}$ is the beamforming coefficient utilized in such a SISO system and $w_{l,k}$ is the weight value in the receiver. After that, the equivalence of the problems \eqref{Prob:SumRate} and \eqref{Prob:WMMSEv1} is obtained by finding the optimal solution of $u_{l,k}$ and $w_{l,k},\forall l,k,$ given the other optimization variables. The detailed proof is omitted due to space limitations. 
\end{IEEEproof}

The new problem formulation in Theorem~\ref{Theorem:WMMSE} is still non-convex, but it has an important desired property: \eqref{Prob:WMMSEv2} is convex with respect to each of the variable sets $\{ u_{l,k}\}$, $\{w_{l,k} \}$, $\{\hat{\rho}_{l,k} \}$, and $\{\rho_{l,k} \}$ when the other three variable sets are treated as constants. In fact, we can find closed-form solutions by equating the first derivatives to zero. We exploit this property to derive an iterative algorithm to find a local optimum (stationary point) to \eqref{Prob:WMMSEv2} in the following subsection.

\begin{figure*}[t]
	\centering
	\includegraphics[trim=0.5cm 7.2cm 22cm 0.38cm, clip=true, width=5.3in]{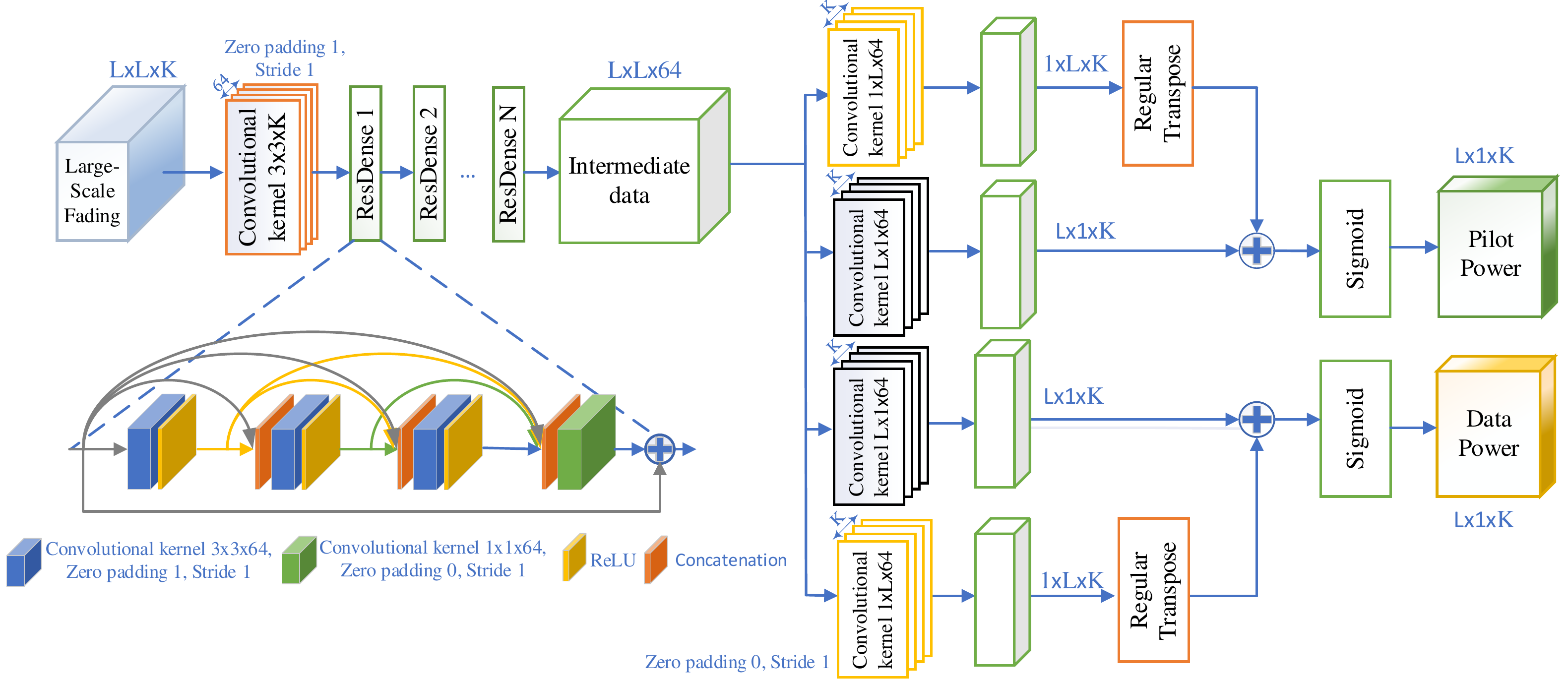} \vspace*{-0.2cm}
	\caption{The proposed PowerNet for the joint pilot and data power control from a given set of large-scale fading coefficients. }
	\label{FigCNN}
	\vspace*{-0.5cm}
\end{figure*} 

\subsection{Iterative Algorithm}
The next theorem provides an iterative algorithm to obtain a stationary point to problem \eqref{Prob:WMMSEv1} by alternatingly updating the optimization variables.

\begin{theorem} \label{Theorem:IterativeAl}
From an initial point $\{ \hat{\rho}_{l,k}^{(0)}, \rho_{l,k}^{(0)} \}$ satisfying the constraints, a stationary point to problem~\eqref{Prob:WMMSEv1} is obtained by updating $\{ u_{l,k}, w_{l,k}, \hat{\rho}_{l,k}, \rho_{l,k} \}$ in an iterative manner. At iteration~$n$, the variables $\{ u_{l,k}, w_{l,k}, \hat{\rho}_{l,k}, \rho_{l,k} \}$ are updated as follows:
\begin{enumerate}[leftmargin=*]
	\item The variables $u_{l,k}$, for all $l,k$, are updated as
	\begin{equation}  \label{eq:ulk}
	u_{l,k}^{(n)} = \sqrt{M  K} \rho_{l,k}^{(n-1)} \hat{\rho}_{l,k}^{(n-1)} \beta_{l,k}^l / \tilde{u}_{l,k}^{(n-1)},
	\end{equation}
	where $\tilde{u}_{l,k}^{(n-1)}$ is given by
	\begin{equation} \label{eq:tildeulk}
	\begin{split}
	 & M K \sum\limits_{i=1}^L (\rho_{i,k}^{(n-1)})^2 (\hat{\rho}_{i,k}^{(n-1)})^2  (\beta_{i,k}^l)^2 +  \left( \sum\limits_{i=1}^L (\hat{\rho}_{i,k}^{(n-1)})^2  \right. \\ 
	 &\times K \beta_{i,k}^l + \sigma_{\mathrm{UL}}^2  \Bigg) \left( \sum\limits_{i=1}^L \sum\limits_{t=1}^K (\rho_{i,t}^{(n-1)})^2 \beta_{i,t}^l + \sigma_{\mathrm{UL}}^2 \right).
	 \end{split}
	\end{equation}
	\item The variables $w_{l,k}$, for all $l,k$, are updated as
	\begin{equation} \label{eq:wlk}
	w_{l,k}^{(n)} = 1/e_{l,k}^{(n)},
	\end{equation}
	where $e_{l,k}^{(n)}$ is
	\begin{equation} \label{eq:elkn}
	e_{l,k}^{(n)} = (u_{l,k}^{(n)})^2 \tilde{u}_{l,k}^{(n-1)}  - 2 \sqrt{M  K} \rho_{l,k}^{(n-1)} \hat{\rho}_{l,k}^{(n-1)} u_{l,k}^{(n)} \beta_{l,k}^l +1.
	\end{equation}
	\item The variables $\hat{\rho}_{l,k}$, for all $l,k$, are updated as in \eqref{eq:rhohatlk}.
	\item The variables $\rho_{l,k}$, for all $l,k$, are updated as in \eqref{eq:rholk}.
\end{enumerate}
 This iterative process converges to a stationary point $\{ u_{l,k}^{\mathrm{opt}}, w_{l,k}^{\mathrm{opt}}, \hat{\rho}_{l,k}^{\mathrm{opt}}, \rho_{l,k}^{\mathrm{opt}}\}$ to problem \eqref{Prob:WMMSEv1} and  $\{ (\hat{\rho}_{l,k}^{\mathrm{opt}})^2, (\rho_{l,k}^{\mathrm{opt}})^2 \}$ is also a stationary point to problem \eqref{Prob:SumRate}. 
\end{theorem}
\begin{IEEEproof}
The closed-form expression for each optimization variable, as shown in \eqref{eq:ulk}--\eqref{eq:rholk}. The expressions for $w_{l,k}$ and $u_{l,k}$ are obtained by taking the first derivative of the objective function due to the fact that there are no constraints on $w_{l,k}$ and $u_{l,k}$. The expressions for the power variables are obtained by taking the first derivative of the Lagrangian function of \eqref{Prob:WMMSEv1} and equating it to zero.
 The convergence of the iterative process in Theorem~\ref{Theorem:IterativeAl} to a stationary point follows from the convexity of the Lagrangian with respect to each of the four sets of optimization variables when the other are constant \cite{Weeraddana2012a}. The solution $\{ \hat{\rho}_{l,k}^{\mathrm{opt}}, \rho_{l,k}^{\mathrm{opt}}\}$ obtained after convergence is a stationary point to~\eqref{Prob:WMMSEv1} and \eqref{Prob:SumRate} as a consequence of the chain rule \cite{Chien2018a}. The detail proof is omitted due to space limitations.
\end{IEEEproof}
Theorem~\ref{Theorem:IterativeAl} provides an iterative algorithm to obtain a local optimum with relatively low computational complexity because each subproblem is solved in closed-form. 
From any feasible initial set of powers $\{ \hat{\rho}_{l,k}^{(0)}, \rho_{l,k}^{(0)} \}$, in each iteration, we update each optimization variable according to \eqref{eq:ulk}--\eqref{eq:rholk} and improve the objective function in each step. This iterative process will be terminated when the variation between two consecutive iterations is small. For instance the stopping condition may be defined for a given accuracy $\epsilon > 0$ as
\setcounter{eqnback}{\value{equation}} \setcounter{equation}{16}
\begin{equation} \label{eq:Stoping}
\left| \sum_{l=1}^L \sum_{k=1}^K R_{l,k}^{(n)} - \sum_{l=1}^L \sum_{k=1}^K R_{l,k}^{(n-1)} \right| \leq \epsilon.
\end{equation}
From Theorem~\ref{Theorem:IterativeAl}, we further observe the relationship of data and pilot power when a user is out of service as the following.
\begin{remark} \label{Remark:phatlkandplk}
In order to maximizing the sum SE, the system may reject some users from service. A particular user~$t$ in cell~$i$ is not served if $\hat{p}_{i,t}^{\mathrm{opt}} = 0$ and $p_{i,t}^{\mathrm{opt}} = 0$. Hence, this user is neither transmitting in the pilot nor data phase.
\end{remark}

\section{A Low-Complexity Implementation Using a Convolutional Neural Network} \label{Sec:CNN}

In this section, we explore the feasibility of using a CNN to learn how to perform joint optimal pilot and data power control, in order to achieve an extremely low-complexity implementation. The input to the CNN is only the large-scale fading coefficients and the output is the data and pilot powers. This is fundamentally different from the previous works \cite{sun2017learning,zappone2018model} that use deep learning to predict the data power allocation based on perfect instantaneous CSI (i.e., small-scale fading).

\subsection{Convolutional Neural Network Architecture}

As the second main contribution of this paper, we introduce a deep learning framework for power allocation in cellular Massive MIMO systems, which uses supervised learning to mimic the power control solution developed in Section~\ref{section:joint-optimization}. We stress that for non-convex optimization problems, a supervised learning with high prediction accuracy provides a good baseline for any further activities, e.g., supervised learning as a warm start for unsupervised learning, to improve the performance of the testing phase \cite{lee2018deep}.

Among the neural network structures in the literature, CNN is currently the most popular family since it achieves higher performance than conventional fully-connected feed-forward deep neural networks. 
We will use a CNN to learn how to perform power control in a given setup, where the BSs locations are fixed and the active users change over time. 
Before going further, we need to make an assumption on how the user locations and large-scale fading coefficients are generated in different realizations of the network.
\begin{assumption} \label{Proposition:UserLocation}
The location of all users are drawn as independent and identically distributed realizations from a given user location distribution, from which the large-scale fading coefficients are also obtained.
\end{assumption}
This assumption indicates that a user should be handled equally irrespective of which number that it has in the cell. A CNN can utilize this property to construct a unified structure for all training samples and reduce the number of parameters as compared to a conventional fully-connected network.\footnote{In this paper, the main task is to design an efficient CNN which is able to obtain performance close to the stationary point with lower runtime than the iterative algorithm in Theorem~\ref{Theorem:IterativeAl}. This CNN may not have the minimal number of parameters. The CNN with lowest cost is left for future work.} The  proposed deep CNN is named PowerNet and is designed to provide good  power control solutions in multi-cell Massive MIMO systems.  In particular, we define a tensor $\mathsf{I} \in \mathbb{R}_{+}^{L\times L \times K}$ containing all the large-scale fading coefficients. We let $\mathsf{O}_d^{\textrm{opt}} \in \mathbb{R}_{+}^{L\times 1 \times  K}$ denote the tensor with the optimal data powers  and  $\mathsf{O}_p^{\textrm{opt}} \in \mathbb{R}_{+}^{L\times 1 \times K}$ denote the tensor with the optimal pilot powers. PowerNet is used to  learn the mapping
\begin{equation} 
\mathcal{F}(\mathsf{I}, \mathsf{O}_d^{(0)},  \mathsf{O}_p^{(0)}  ) = \{ \mathsf{O}_d^{\textrm{opt}} , \mathsf{O}_p^{\textrm{opt}} \}, 
\end{equation}
where $\mathsf{O}_d^{(0)}$ and $\mathsf{O}_p^{(0)}$ are the initial set of data and pilot powers, respectively. $\mathcal{F}(\cdot, \cdot, \cdot)$ is the continuous process to obtain the optimal powers from the input set of large-scale fading coefficients and the initial power tensors $\mathsf{O}_d^{(0)}, \mathsf{O}_p^{(0)}$. \footnote{The proposed neural network involves a series of activities such as convolutional operations and ReLUs adopted to predict pilot and data power variables to the sum SE problem as demonstrated Fig.~\ref{FigCNN}. Comprehensive mathematical explanation step by step over modules is omitted due to space limitations.} We adopt the state-of-the-art residual dense block (ResDense) \cite{zhang2018a} that consists of densely connected convolutions \cite{huang2017} with the residual learning \cite{he2016deep}. As shown in Fig.~\ref{FigCNN}, a ResDense block inherits the Densely Connected block in \cite{huang2017} with residual connection similar to \cite{he2016deep}. Compared with ResDense in \cite{zhang2018a}, we have an additional (rectified linear unit) ReLU activation after the residual connection. 

Our proposed PowerNet is constructed from $N$ sequentially connected ResDense blocks to extract better features from the large-scale fading coefficients (we observed that $N=5$ is sufficient for our problem to strike a balance between prediction accuracy and computational complexity). The input and output size of the neural network are different. We are therefore using multiple 1D convolutions to make the sides equal. To exploit the correlation in both horizontal and vertical direction, both horizontal and vertical 1D convolutions are used. A regular transpose layer is applied following vertical 1D convolution to ensure a data size of $L \times 1 \times K$. The output of these two 1D convolutions are summed up to obtain the final prediction output. This prediction is used for both pilot and data power as depicted in Fig.~\ref{FigCNN}. When training PowerNet, we use a loss function based on the Frobenius norm as
\begin{equation} \label{eq:Loss}
\begin{split}
f(\Theta) =&   \|  \mathsf{O}_{d} - \mathsf{O}_{d}^{\mathrm{opt}}  \|_F^2  +  \| \mathsf{O}_{p} - \mathsf{O}_{p}^{\mathrm{opt}} \|_F^2,
\end{split}
\end{equation}
where $\Theta$ includes all convolution kernels and biases used in our neural network.
The loss in \eqref{eq:Loss} is applied for one realization of user locations, so the ultimate loss is averaged over the training data set.

\subsection{Data Set \& Training Process}
In order to train the PowerNet, we generated a training set with more than $1 000 000$ realizations of user locations (i.e., large-scale fading coefficients) and the corresponding output $\mathsf{O}_{p}^{\mathrm{opt}}, \mathsf{O}_{d}^{\mathrm{opt}}$ obtained by our new algorithm presented in Theorem~\ref{Theorem:IterativeAl}. We generated $3000$ and $100$ mini-batches of size $L \times L \times K \times 512$ for the training and testing phase, respectively. The number of epochs was selected to be a function of the network size, e.g., it is $200$ epochs if $L=4, K=5$ and $1000$ epochs if $L=9,K=10$. We use a momentum of $0.99$ and babysitting of the learning rate (varing from $10^{-3}$ to $10^{-5}$) to get the best prediction performance and minimize the training time as well. We note that the learning rate may be reduced by approximately $3$ times if the test loss remains the same for $50$ consecutive epochs. The Adam optimization \cite{kingma2014adam} was used to train PowerNet.

\begin{figure*}[t]
	\noindent\begin{minipage}{0.32\textwidth}
		\centering
		\includegraphics[trim=0.55cm 0cm 0.3cm 0cm, clip=true, width=2.45in]{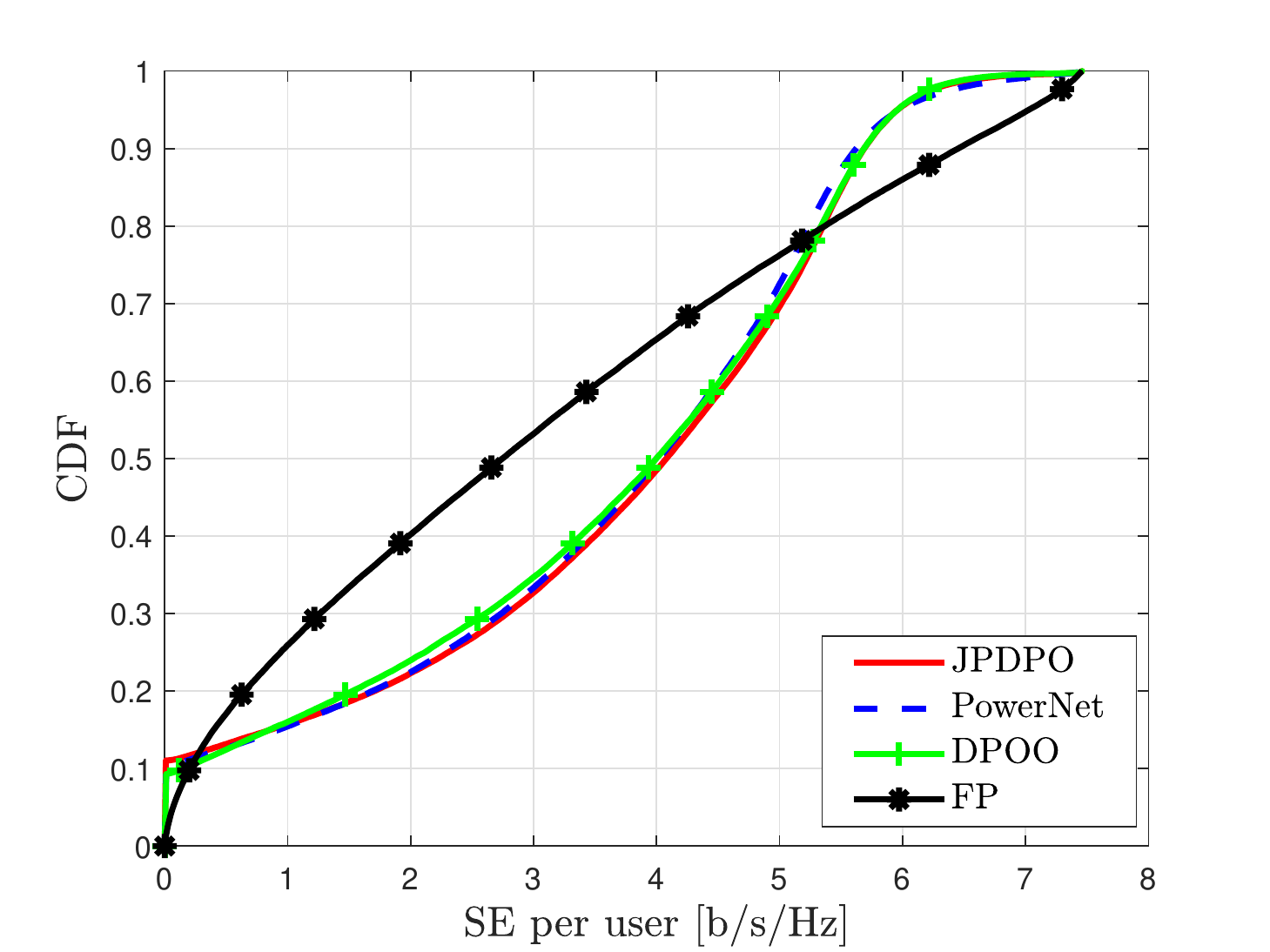} \vspace*{-0.5cm}
		\caption{CDF of SE per user [b/s/Hz] with $L=4$ and $K = 5$.}
		\label{FigPower}
		\vspace*{-0.45cm}
	\end{minipage}
	\hfill
	\noindent\begin{minipage}{0.32\textwidth}
		\centering
		\includegraphics[trim=0.55cm 0cm 0.3cm 0cm, clip=true, width=2.4in]{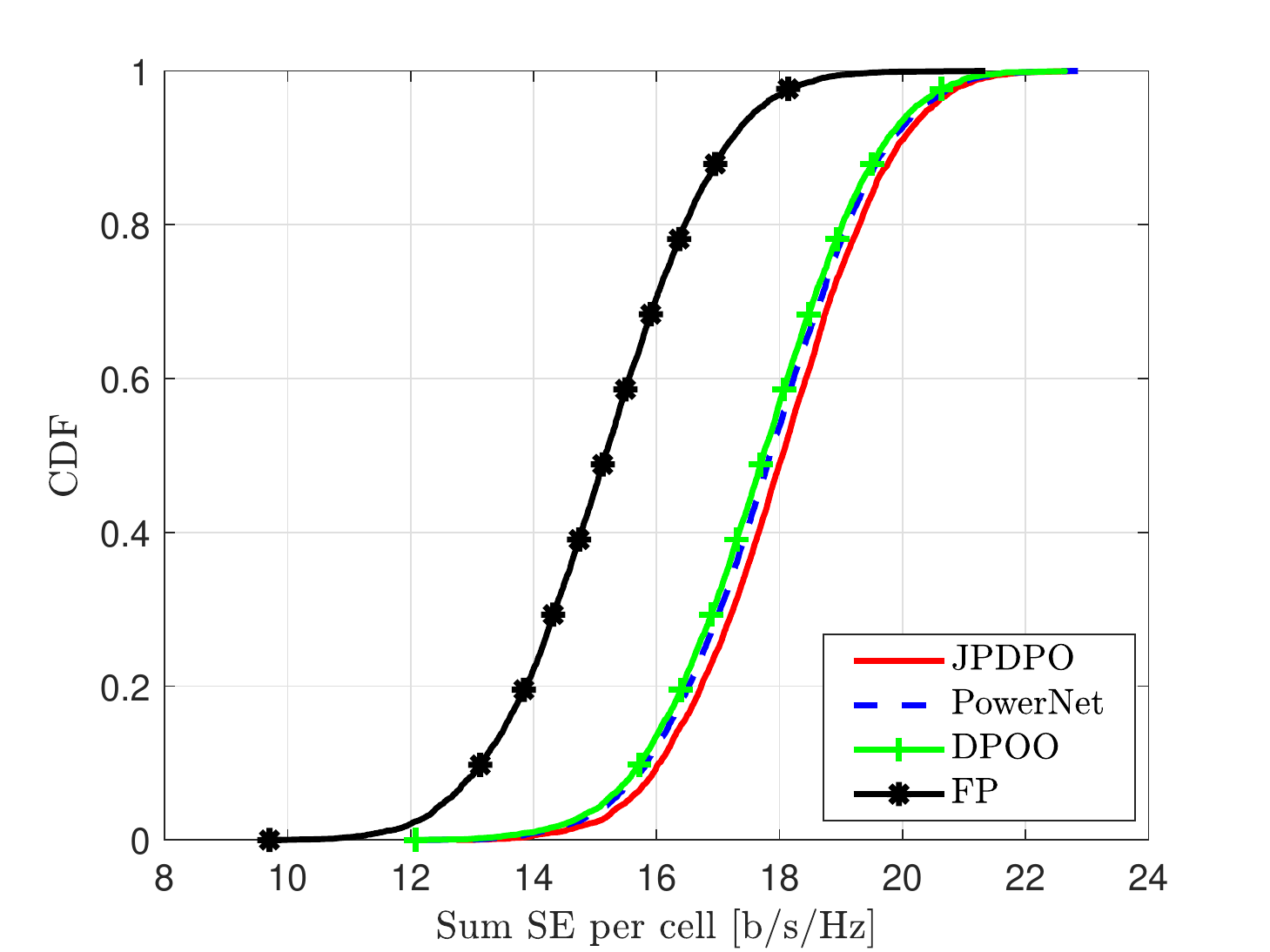} \vspace*{-0.5cm}
		\caption{CDF of sum SE per cell [b/s/Hz] with $L=4$ and $K = 5$.}
		\label{FigCDFL4K5}
		\vspace*{-0.45cm}
		\noindent \end{minipage}
	\hfill
	\begin{minipage}{0.32\textwidth}
		\centering
		\includegraphics[trim=0.55cm 0cm 0.3cm 0cm, clip=true, width=2.4in]{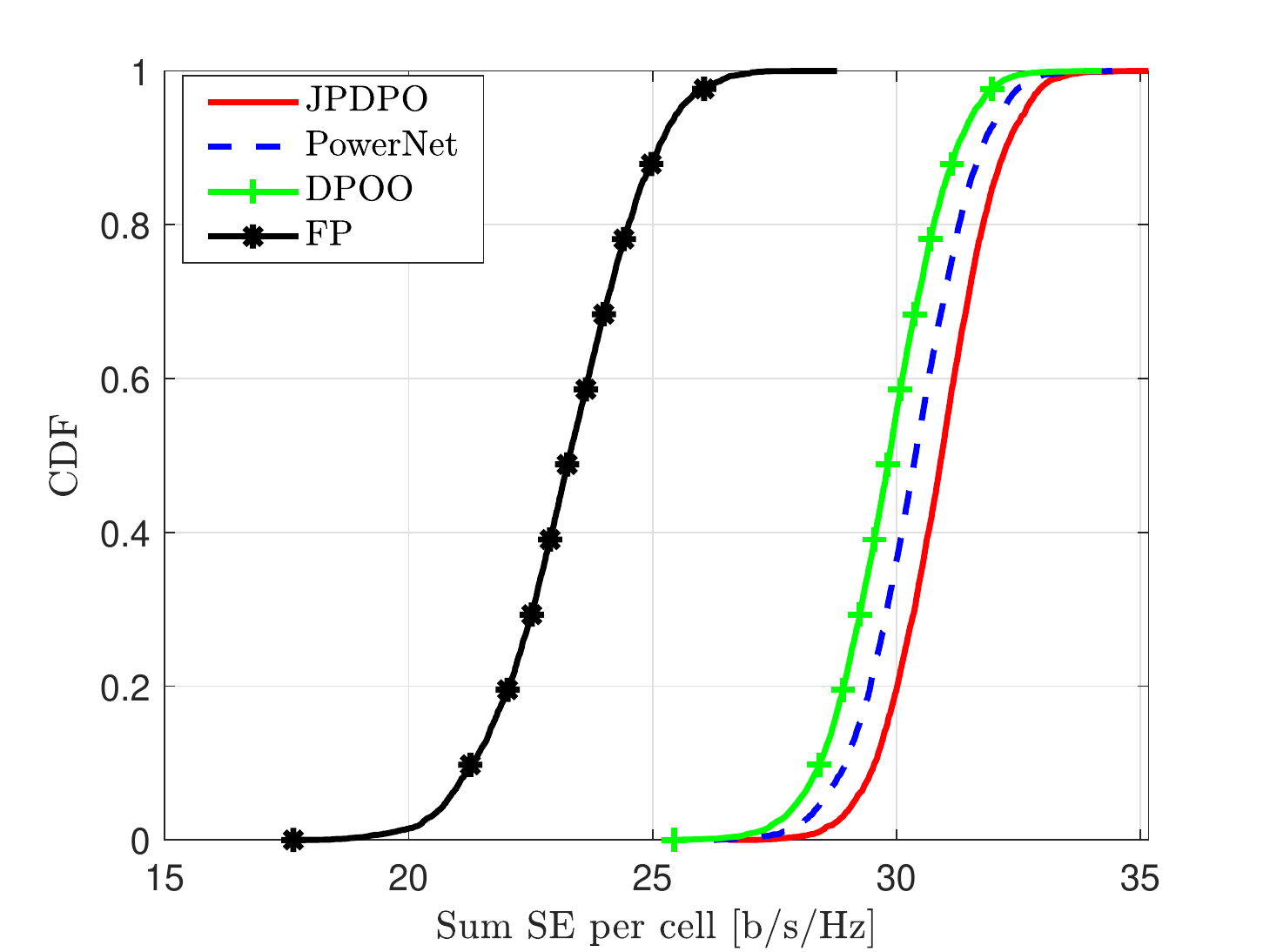} \vspace*{-0.5cm}
		\caption{CDF of sum SE per cell [b/s/Hz] with $L=9$ and $K = 10$.}
		\label{FigCDFL9K10}
		\vspace*{-0.45cm}
	\end{minipage}
\end{figure*}
  
\section{Numerical Results}
For simulations, we consider a multi-cell Massive MIMO system with $L$ square cells ($L =4, 9$) in a square area of $1$~km$^2$. In every cell, the BS is equipped with $200$ antennas and located at the center. $K$ users are uniformly distributed in each cell at distances to the BS that are larger than $35$\,m. To simulate interference in real cellular systems, the coverage area is wrapped around. The system uses a bandwidth of $20$ MHz and the noise power is $-96$ dBm.  Each coherence interval consists of $200$ symbols. The $3$GPP pathloss model from \cite{LTE2010b} is used to compute the large-scale fading coefficients as
\begin{equation}
\beta_{i,t}^l \mbox{[dBm]} = -148.1 - 37.6\log_{10} (d_{i,t}^l / 1\,\mbox{km})  + z_{i,t}^l,
\end{equation}
where $d_{i,t}^l$ is the distance between user~$t$ in cell~$i$ and BS~$l$, while $ z_{i,t}^l$ is the related shadow fading that follows a Gaussian distribution with zero mean and standard derivation $7$~dB. The maximum power level is $P=200$~mW. Simulation results of the following benchmarks are considered for comparison: 
\begin{itemize}[leftmargin=4mm]
	\item[$1)$] \textit{Fixed power level}: The system uses an equal power level $200$ mW to allocate for both pilot and data symbols. It is denoted as ``FP'' in the figures. 
	\item[$2)$] \textit{Data power optimization only}: The system uses a simplification of Theorem~\ref{Theorem:IterativeAl} to perform data power control, while the pilot power is fixed at the maximum of $200$ mW. It is denoted as ``DPOO'' in the figures. 
	\item[$3)$] \textit{Joint pilot and data power optimization}: The system uses Theorem~\ref{Theorem:IterativeAl} to find the optimal pilot and data powers for all users. It is denoted as ``JPDPO'' in the figures.
	\item[$4)$] \textit{Joint pilot and data power optimization based on deep learning}: The system uses the proposed CNN described in Section \ref{Sec:CNN} to find the pilot and data powers for all users. It is denoted as ``PowerNet'' in the figures. 
\end{itemize}
\subsection{Power Prediction Performance \& Sum Spectral Efficiency}
The accuracy of the proposed neural network in predicting the SE per user [b/s/Hz] is shown in Fig.~\ref{FigPower} for a multi-cell Massive MIMO system with $L=4$ and $K=5$. The predicted SEs produced by PowerNet are almost equal to those obtained by Theorem~\ref{Theorem:IterativeAl}. The average prediction error is $1\%$.
 Fig.~\ref{FigCDFL4K5} displays the cumulative distribution function (CDF) of the sum SE per cell [b/s/Hz] in the same setup as in Fig.~\ref{FigPower}. Because of the high prediction accuracy for the power coefficients, the sum SE per cell obtained by PowerNet is only about $1\%$ lower than what is obtained by using Theorem~\ref{Theorem:IterativeAl} directly. We observe that FP performs much worse than the case with optimized power, while DPOO with only optimized data powers achieve $98\%$ of the sum SE that is produced by the joint pilot and data power control. Fig.~\ref{FigCDFL9K10} considers a larger network with $9$ cells serving $90$ users in total. The gap is now even larger between FP and the cases with optimized powers. In particular, by jointly optimizing the pilot and data power, $33\%$ higher sum SE is achieved. Applying power control for both the data and pilot powers give about $4\%$  higher sum SE than the counterpart that only optimizes the data power. PowerNet achieves $98\%$ of Theorem~\ref{Theorem:IterativeAl}.
\subsection{Runtime}
\begin{table}[t]
\caption{Average runtime of different methods in millisecond.}
	\centerline{ 
			\begin{tabular}{|c|c|c|c|c|}
				\hline
				\diagbox[width=10em]{System \\ specifications}{Benchmark} &JPDPO & DPOO & \thead{ PowerNet \\ (CPU) }& \thead{PowerNet \\ (GPU)} \\
				\hline
				$L=4,$ $ K=5$ & $42.24$  & $30.90$ & $2.99$ & $0.0177$ \\
				\hline
				$L=9 , K=10$ & $491.08$ & $269.33$ & $14.90$ & $0.0283$ \\
				\hline				
			\end{tabular}
		} \vspace*{-0.5cm}
		\label{tablerunningitme}
\end{table}
To numerically evaluate the computational complexity of all benchmarks, we implement the testing phase using MatConvNet \cite{vedaldi2015matconvnet} on a Windows $10$ computer, with the central processing unit (CPU) AMD Ryzen $1950$x $16$-Core, 3.40 GHz, and a graphics processing unit (GPU) Titan XP Nvidia GPU. The proposed neural network is tested when using only the CPU and when using both CPU and GPU.
For comparison, Theorem~\ref{Theorem:IterativeAl} was also implemented in MATLAB.

The average runtimes are given in Table~\ref{tablerunningitme}, where the accuracy value $\epsilon$ in \eqref{eq:Stoping} is $0.01$. We implemented Theorem~\ref{Theorem:IterativeAl} in a sequential fashion, but the runtime is anyway low. For a system with $4$ cells, each serving $5$ users, the runtime for Theorem~\ref{Theorem:IterativeAl} is $42.24$~ms, while it is $30.90$~ms if we only consider data powers as optimization variables. If there are $9$ cells and $10$ users per cell, the runtime increased by $12 \times$ and $8.72 \times$, respectively.
These numbers can potentially be reduced by a factor $LK$ with an ideal parallelized implementation.

When PowerNet runs on the CPU only, the runtime is  $2.99$~ms with $L=4, K=5$, and $14.90$~ms with $L=9, K = 10$. When using the GPU, the runtime of PowerNet is reduced to $0.0177$~ms and $0.0283$~ms, respectively. Since the typical duration of a coherence interval is around  $1$~ms, this means that PowerNet implemented on a GPU can be used for real-time power control in multi-cell Massive MIMO systems.

\section{Conclusion}
This paper has investigated the joint pilot and data power control for the sum SE maximization in uplink Massive MIMO systems. This is a non-convex problem but we proposed a new algorithm, inspired by weighted MMSE approach, to find a stationary point. The joint pilot and data power optimization can obtain $30\%$ higher sum SE than equal power transmission. We used the proposed algorithm to also construct a neural network, called PowerNet, that predicts both data and pilot powers very well, leading to only about $2\%$ loss in sum SE in a multi-cell system serving $90$ users. PowerNet uses only the large-scale fading coefficients to predict the power control, making it scalable to Massive MIMO systems with many antennas.
It has a runtime that is far below a 1\,ms, meaning that it enables real-time power control in systems where new power control coefficients need to be obtained at the millisecond level due to changes in the scheduling decisions or user mobility. This demonstrates the feasibility of using deep learning for real-time power control in Massive MIMO, but we stress that other methods can potentially reach the same or better runtime.

\bibliographystyle{IEEEtran}
\bibliography{IEEEabrv,refs}
\end{document}